\documentclass{pnastwo}
\usepackage[pdftex]{graphicx}

\usepackage{amssymb,amsfonts,amsmath}

\begin{document}

\title{The noisy edge of traveling waves}

\author{Oskar Hallatschek\affil{1}{Biophysics and Evolutionary
    Dynamics Group, Max Planck Institute for Dynamics \&
    Self-Organization, G\"ottingen, Germany}}

\contributor{Submitted to Proceedings of the National Academy of Sciences
of the United States of America}

\maketitle

\begin{article}

\begin{abstract}
  Traveling waves are ubiquitous in nature, and control the speed of
  many important dynamical processes, including chemical reactions,
  epidemic outbreaks and biological evolution. Despite their
  fundamental role in complex systems, traveling waves remain elusive
  because they are often dominated by rare fluctuations in the wave
  tip, which have defied any rigorous analysis so far.  Here, we show
  that by adjusting nonlinear model details, noisy traveling waves
  can be solved exactly. The moment equations of these tuned models
  are closed and have a simple analytical structure resembling the
  deterministic approximation supplemented by a non-local cutoff
  term. The peculiar form of the cutoff shapes the noisy edge of
  traveling waves and is critical for the correct prediction of the
  wave speed and its fluctuations. Our approach is illustrated and
  benchmarked using the example of fitness waves arising in simple
  models of microbial evolution, which are highly sensitive to number
  fluctuations.  We demonstrate explicitly how these models can be
  tuned to account for finite population sizes, and determine how
  quickly populations adapt as a function of population size and
  mutation rates. More generally, our method is shown to apply to a
  broad class of models, in which number fluctuations are generated by
  branching processes. Due to this versatility, the method of model
  tuning may serve as a promising route towards unraveling universal
  properties of complex discrete particle systems.
\end{abstract}

\dropcap{T}he wave-like spread of discrete entities pervades our
everyday life. For example, the spread of ions, pathogens and
beneficial mutations control the human heart beat, the yearly threat
of influenza and evolutionary progress~\cite{murray-book}. A thorough
understanding of how these waves form and spread has numerous
applications ranging from the control of chemical
reactions~\cite{winfree1972spiral,kuramoto2003chemical} to the
prediction of epidemic
outbreaks~\cite{grenfell2001travelling,brockmann2006scaling}. Great
research effort has therefore been made on the question of how
traveling waves emerge in complex systems from the multiplicity of
relatively simple interactions, in particular the random dispersal of
particles and reactions between particles. To simplify the analysis,
most theoretical studies have been neglecting number fluctuations,
which are inevitable in systems of discrete particles. However, when
stochastic simulations became feasible, it was found that those
previously ignored fluctuations can have a strong impact on the
dynamics of
waves~\cite{Tsimring:1996p13322,Brunet:VelocityShiftCuttoff,saarloos03}.
Simple models of biological evolution (described below) serve as
the prime example for this drastic sensitivity on noise because they
are dominated by the few most fit individuals in a population, and
break down if number fluctuations are
neglected~\cite{Tsimring:1996p13322}. With the added difficulty of
particle discreteness, the analysis of traveling waves became one of
the important challenges of statistical physics and mathematical
biology, which still defies systematic analytical techniques.

Here, we show that an exact analysis of noisy traveling waves is
feasible when the nonlinear details of the underlying model are
chosen appropriately. The main idea is to \emph{tune} the
nonlinearities to obtain the least difficult math, while retaining
the universal features of the model. Specifically, we show that it is
possible to close the hierarchy of moment equations using a suitably
designed nonlinear constraint on the dynamics. The tuned model can
then be used to extract the universal features of noisy traveling
waves, for which only heuristic approaches have been available so
far. The method of model tuning can be naturally generalized to a wide
range of unsolved stochastic nonlinear problems, and therefore
provides a promising new tool to unraveling universal features of
non-equilibrium systems.

\section{Noisy wave models}
\label{sec:model-1}

\begin{figure}
\centerline{  \includegraphics[width=.4\textwidth]{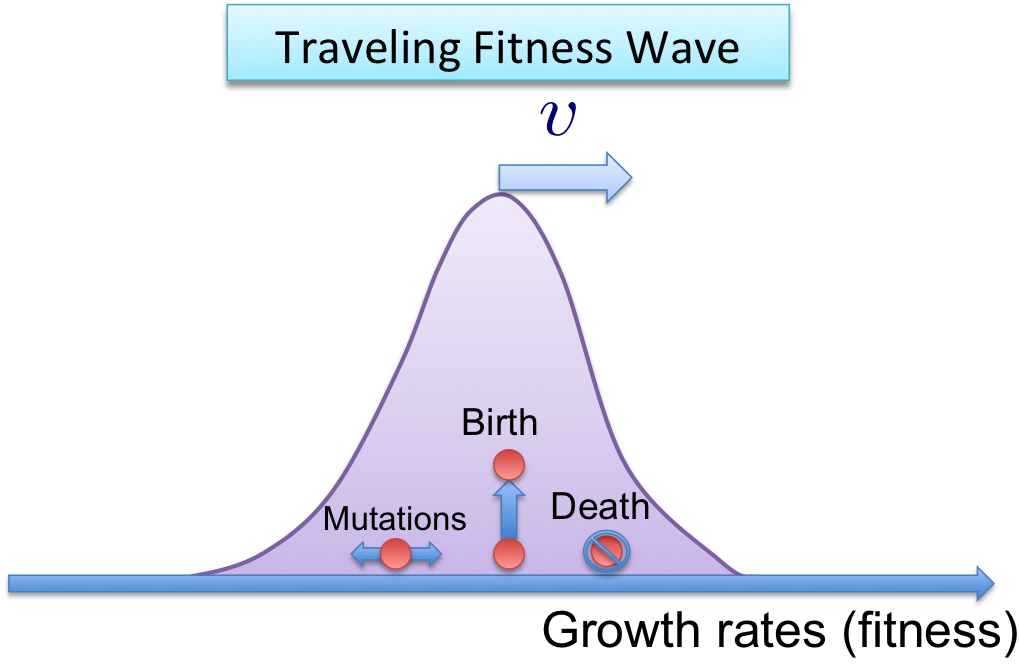}}
  \caption{ \label{fig:cover} A paradigmatic example for noisy traveling
  waves are ``fitness'' waves arising in simple models of
  evolution. The colored particles represent individuals with
  characteristic growth rates, or fitnesses (horizontal
  axis). Individuals can mutate, replicate (``birth'') and be
  eliminated from the gene pool (``death''), as illustrated. These
  simple dynamical rules give rise to a distribution of growth rates
  resembling a bell-like curve at steady state, which propagates
  towards higher growth rates like a solitary wave. The random fluctuations
  in the tails of the wave have precluded any rigorous analysis in the
  past. }
\end{figure}

Simple models of
evolution~\cite{Tsimring:1996p13322,Peng:2003p14364,Snyder:2003p14532,Kloster:2004p14395,Derrida:2007p1381,desai-fisher-genetics-2007}
provide spectacular examples of noisy traveling waves because of their
drastic sensitivity to rare fluctuations in the wave tip.  As
illustrated in figure \ref{fig:cover}, these waves describe the
continual increase of growth rate, also called fitness, due to
spontaneous mutations in a finite population. The wave speed, which is
a measure for how quickly populations adapt, increases without bound
if number fluctuations are neglected~\cite{Tsimring:1996p13322}. To
reproduce a \emph{steady} state with constant wave speed observed in
stochastic simulations, any analysis has to incorporate number
fluctuations, at least heuristically by introducing an ad hoc
``cutoff'' in the noisy tip of the
wave~\cite{Tsimring:1996p13322}. The relationship between traveling
speed and population size has been investigated extensively in the
recent
literature~\cite{Rouzine:2003p13321,park2007clonal,desai-fisher-genetics-2007,rouzine2008traveling},
with some controversy regarding the universal
behavior~\cite{Brunet:2008p14573,park2010speed}, in order to interpret growth rate
measurements in microbial evolution
experiments~\cite{gerrish1998fate,desai2007speed}.

We now take a closer look at the models underlying these fitness waves
to elucidate their characteristic structure. This will lead us to a
general model of noisy traveling waves, which forms the basis of our
analytic approach. Most evolution models implement the reproduction of
individuals by a standard branching process. The growth rate is
subject to small variations due to random mutations, which are modeled
through a random walk, for instance by standard
diffusion~\cite{Tsimring:1996p13322}. In addition, all evolution
models contain a nonlinearity that limits the total population
size. This accounts for the fact that natural populations must stay
finite due to resource limitations.

\begin{figure}
  \includegraphics[width=.4\textwidth]{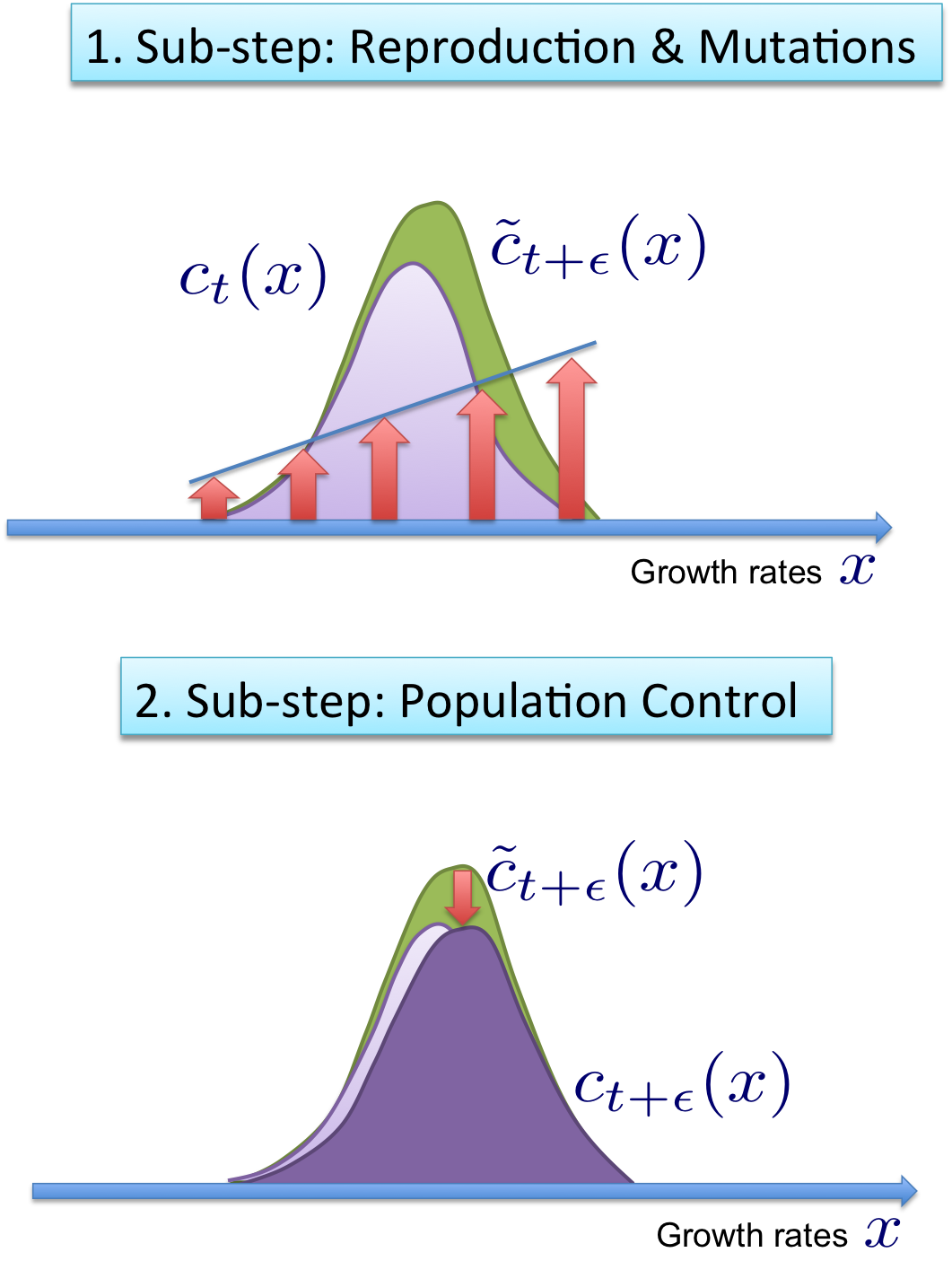}
  \caption{\label{fig:CBRW} A computational timestep in fitness wave
    models consists of two sub-steps. The first sub-step (upper panel)
    is reproduction and mutations. Growth favors the most fit
    individuals as indicated by the red arrows. The resulting
    population density $\tilde c(x)$ will generally
    violate the constrained of constant population size. The second sub-step
    (lower panel) restores the prescribed population size by a random
    elimination of individuals from the population (this step may also
    be interpreted as Darwinian selection). The result of the whole
    timestep is a constrained branching random walk of ``genotypes'',
    which shifts the fitness distribution towards higher fitness. }
\end{figure}

The generic mathematical structure of such evolution models can be
framed as follows. The state of the population at time $t$ is
described by a function $c_t(x)$ that represents the number density of
individuals having a growth rate $x$. The population is assumed to
evolve in discrete timesteps of size $\epsilon$, which is eventually
sent to zero in order to obtain a continuous time Markov process. Any
such timestep consists of two sub-steps. The first one, illustrated
in Fig. \ref{fig:CBRW}A, embodies the mutational process, which leads
to slight variations in growth rates, and the process of reproduction,
which causes individuals with growth rate above the mean to increase
in relative number. In general, the combined effect of both processes
can be described by the equation
\begin{equation}
  \label{eq:BRW-genotype}
  \tilde c_{t+\epsilon}-c_t=\epsilon \mathcal{L}c_t + \sqrt{2\epsilon
    c_t}\,\eta \;,
\end{equation}
which takes the concentration field from $c_t$ to an intermediate
value $\tilde c_{t+\epsilon}$. The term $\epsilon \mathcal{L}c_t$
represents the \emph{expected} change in density due to reproduction
and mutations. This term is linear in the concentration field because
the number of offspring and mutants per timestep is proportional to
the current population density. The linear operator $\mathcal{L}$ is
similar to a Hamiltonian in physics. A specific example will be
discussed later on. The stochastic term $\sqrt{2\epsilon c_t}\eta$ in
equation (\ref{eq:BRW-genotype}) accounts for all random factors that
influence the reproduction process. The function $\eta(x)$ represents
standard white noise, i.e., a field of delta correlated random
numbers, $\overline{ \eta(x)\eta(y)}=\delta(x-y)$, where
$\overline{f}$ denotes the ensemble average of a random variable
$f$. The amplitude $\propto \sqrt{\epsilon c_t}$ of the noise term in
equation (\ref{eq:BRW-genotype}) is typical for number fluctuations:
Due to the law of large number, the expected variance from one
timestep to the next is proportional to the number $\epsilon c_t$ of
expected births or deaths during one timestep.

Because the reproduction step changes particle numbers, another
sub-step of population control is required to enforce a
constant population size. In most models and
experiments~\cite{gerrish1998fate,desai2007speed}, this step is
realized by a random culling of the population: individuals are
eliminated at random from the population until the population size
constraint is restored, see Fig.~\ref{fig:CBRW}B. Mathematically, the
selection step can be cast into the form
\begin{equation}
  \label{eq:random_culling}
  c_{t+\epsilon}=\tilde c_{t+\epsilon}(1-\lambda)\;,
\end{equation}
where $\lambda$ represents the fraction of the population that has to
be removed to comply with the population size constraint. The
second sub-step completes the computational timestep,
and takes the concentration field from the intermediate state $\tilde
c_{t+\epsilon}$ to the properly constrained state $c_{t+\epsilon}$.

The combination of noise and nonlinearity (via the population
constraint) has led many theoretical studies to engage in
approximations that are often hard to justify or to
control\footnote{A remarkable exception is the exactly solvable
  ``exponential'' model by Brunet et
  al.~\cite{brunet2007effect}}. Here, we take a different approach by
optimizing the \emph{form} of the nonlinearities in order to obtain an
exactly solvable model. To this end, we first generalize the above
model in the following way. Instead of enforcing a fixed population size, we
allow for a whole class of constraints. Specifically, we assume that
the selection step enforces a constant value of $1$ for the inner
product of the concentration field $c_t(x)$ and a new function $u(x)$,
\begin{equation}
  \label{eq:7}
  1=\int_x u(x) c_t(x)\equiv \langle u\mid c_t\rangle\;.
\end{equation}
Here, we have introduced the ``bra-ket'' notation commonly used in
quantum mechanics. The function $u(x)$ defines the selection rule and
will be called \emph{selection function} from now on. It will serve
as a ``tuning wheel'' in the following. If one chooses $u=N^{-1}$, one
obviously recovers the fixed population size constraint. For any other
choice, the population size will not be fixed. At best, one obtains a
steady state with a population size fluctuating around its mean value,
$\overline N$.

\section{Results}
\label{sec:results}

\subsection{Model tuning}
\label{sec:model-tuning}
Equations (\ref{eq:BRW-genotype}, \ref{eq:random_culling}, \ref{eq:7})
define a class of models that generate branching random walks subject
to a global constraint. As we argue in the Discussion section, these
features not only characterize fitness waves but furnish the essence
of most noisy traveling waves. The fundamental difficulty in analyzing
such models becomes apparent when we try to determine the typical
dynamics by averaging over the stochastic noise term.  This can be
done by eliminating $\lambda$ using the constraint, equation
(\ref{eq:7}), sending $\epsilon$ to zero and carrying out the ensemble
average. This straight forward calculation, detailed in the
Supporting Information, yields
\begin{equation}
  \label{eq:9}
  \partial_t \overline c_{t}=\left(\mathcal{L}-2 u\right)
  \overline c_t-\overline{\langle c_t\mid \left(\mathcal{L}^\dagger- 2
      u\right)u\rangle
  c_t}
\end{equation}
for the mean concentration field $\overline c_t$. Equation
(\ref{eq:9}) reflects the usual ``horror'' inherent to nonlinear
stochastic problems: Moment equations do not close in general. The
mean depends on the second moment, the second on the third, and so
on. A whole hierarchy of moment equations would have to be solved
self-consistently to make progress.

However, if we consider $u(x)$ as a tunable function, equation (\ref{eq:9})
allows for different kind of simplification. Suppose, the solution
$u_*(x)$ of 
\begin{equation}
  \label{eq:12}
  \left(\mathcal L^\dagger - 2 u_*\right)u_* =0
\end{equation}
exists, and we choose $u_*(x)$ as the selection function. For this
particular model, the nonlinearity in equation (\ref{eq:9}) disappears
identically. Thus, the dynamics of the first moment becomes linear,
\begin{equation}
  \label{eq:20}
  \partial_t \overline c_t=\left(\mathcal L - 2 u_*\right) \overline c_t \;,
\end{equation}
and hence solvable. In particular, if a steady state exists, the
steady state concentration field $\overline c $ is in the null space
of $\mathcal L - 2 u_*$. These observations suggest an algorithm to
construct for a given linear operator $\mathcal L$, a solvable
constrained branching walk model: First, identify the selection
function $u_*(x)$ for which the averaged dynamics becomes simple
(i.e. linear). To this end, one has to solve equation (\ref{eq:12}),
which is in general a deterministic partial differential
equation. Second, solve the corresponding moment equation
(\ref{eq:20}), which is guaranteed to be linear for the chosen
selection function $u_*(x)$.

At this point it is not clear, whether
equations~(\ref{eq:12},~\ref{eq:20}) have stationary solutions, and,
if so, whether these solutions can be used to extract universal
features of traveling waves. Therefore, we apply our recipe to a
simple model of fitness waves. This model is defined by the operator
\begin{equation}
  \label{eq:3}
  \mathcal{L}_{\textnormal{\scriptsize evo}}=(x-x_0 )+v\partial_x+ D\partial_x^2 \;,
\end{equation}
which corresponds to a branching random walk on a reaction rate
gradient~\cite{Cohen:2005p13342}. It has been proposed as the simplest
model of asexual evolution in order to explain the fitness growth
observed in directed evolution experiments with large populations of
RNA viruses~\cite{Tsimring:1996p13322}. The term $(x-x_0) c_t$ in
equation~(\ref{eq:3}) represents the reproduction with $x_0$ being the
mean growth rate of the population. The mutational process is modeled
as a diffusion process with diffusivity $D$. This diffusion
approximation is justified when mutations occur at a higher rate than
the typical growth rate difference they confer, which is an
appropriate assumption for viruses with large mutation rates.  For
microbes with lower mutation rates, it is necessary to discretize the
fitness landscape and model mutations as fitness jumps of finite
size. The discretized scenario is slightly more complex as it has one
additional parameter, namely the characteristic fitness effect of
mutations, but can be discussed in complete analogy to the present
case of a continuous fitness landscape (Supporting Information).
Finally, the term $v\partial_x c_t$ appears in equation~(\ref{eq:3})
because the dynamics of the fitness wave is described in a reference
frame moving with the average speed $v$ of the fitness wave.

Using $\mathcal{L}_{\textnormal{\scriptsize evo}}$ in
equation~(\ref{eq:12}), we find that the equation
\begin{equation}
  \label{eq:22}
  D\partial_x^2u_*-v\partial_x u_*+xu_*- 2 u_*^2=0 
\end{equation}
specifies the selection function $u_*(x)$. Once $u_*$ is determined,
the mean concentration field follows as the stationary solution of the
linear equation~(\ref{eq:20}). This second task requires no special
effort if the random walks are simple diffusion processes as in
equation~(\ref{eq:3}). Then, $\overline c$ at steady state and $u_*$ are
simply related by
\begin{equation}
  \label{eq:24}
  \overline c\propto u_* e^{-v x/D}\;,
\end{equation}
which generates equation~(\ref{eq:22}) when inserted into
equation~(\ref{eq:20}). Thus, $\overline c$ follows from $u_*$
directly by multiplying with a decaying exponential. The prefactor in
equation~(\ref{eq:24}) is set by the constraint equation~(\ref{eq:7}).

\begin{figure}
  \includegraphics[width=.45\textwidth]{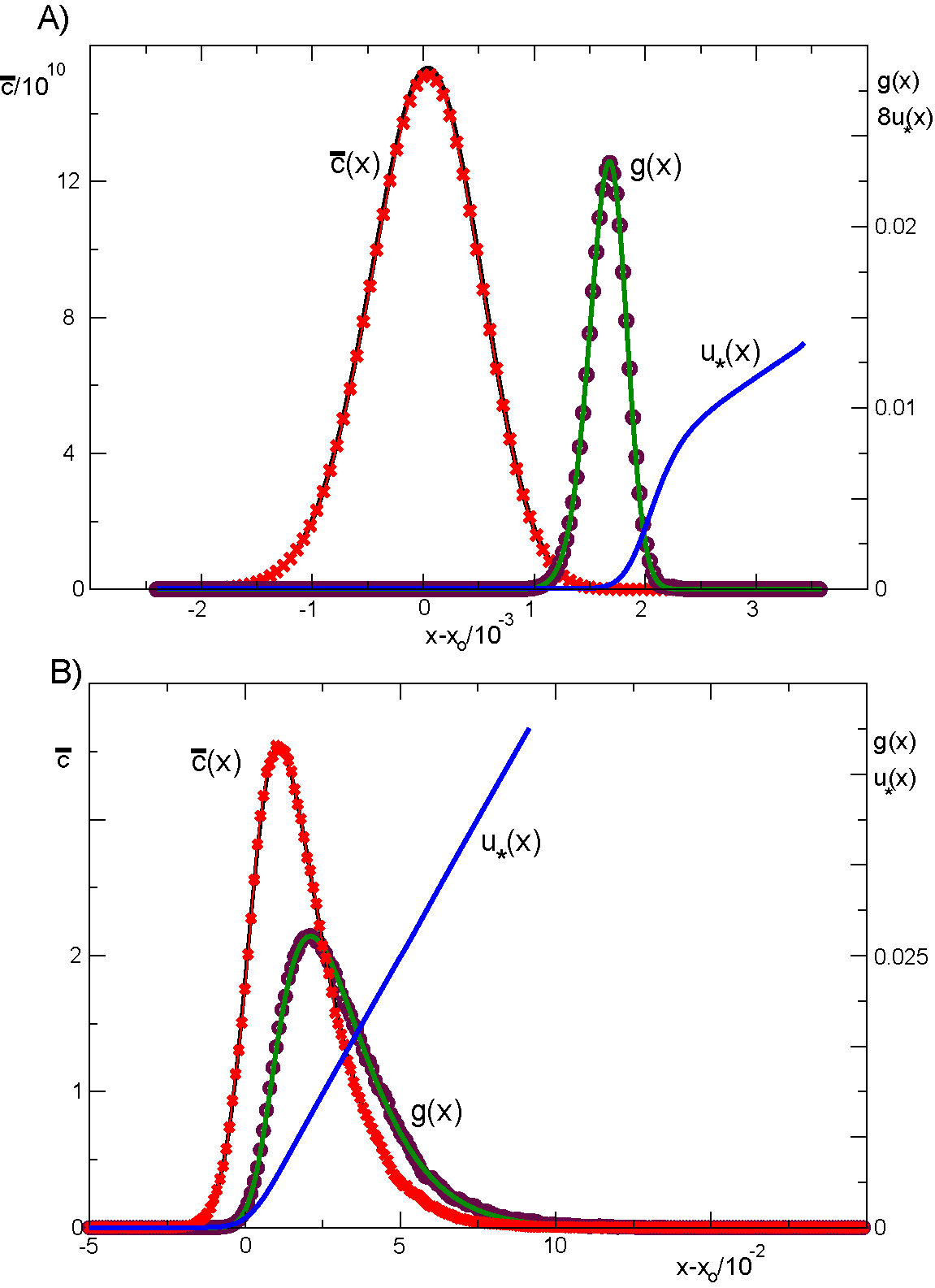}
  \caption{\label{fig:c-u-g} Stochastic simulations confirm exact
    theoretical predictions. The mean population density $\overline c$
    and the functions $u_*(x)$, $g\equiv u_*\overline c$ for two
    different choices for the parameters of the evolution model are
    depicted. The selection function $u_*(x)$ has the intuitive
    interpretation of a fixation probability, see main text. The
    distribution $g\equiv u_*\overline c$ indicates from which
    co-ordinate the individual is sampled whose descendants will
    eventually take over the front. To obtain the depicted curves, we
    first identified $u_*(x)$ by solving equation~(\ref{eq:22})
    numerically and then determined $\overline c(x)$ by stochastic
    simulations of the tuned model with selection function
    $u(x)=u_*(x)$. Notice the near perfect agreement between stochastic
    simulations (symbols) and theory (solid lines). The control
    parameter for A) and B) was set to $vD^{-2/3}=5.17$ ($D=10^{-11}$)
    and $vD^{-2/3}=.464$ ($D=10^{-7}$), respectively.}
\end{figure}

The set of equations (\ref{eq:22},~\ref{eq:24}) is controlled by
just one parameter $v D^{-2/3}$, as can be seen by introducing scaled
variables. In Fig.~\ref{fig:c-u-g}, we depict representative numerical
solutions for both regimes of large and small values of this control
parameter, and compare with simulations of the tuned model. We find
that both regimes exhibit strongly different wave profiles.  Whereas
the mean concentration $\overline c$ exhibits for the most part a
Gaussian shape in regime of large wave speeds ($v D^{-2/3} > 1$ ), it
is strongly skewed for small speeds.  Notice the nearly perfect
agreement between numerics and stochastic simulations, which confirms
our theoretical framework and demonstrates the feasibility of our
model tuning approach.

\subsection{Interpretation of the tuned models}
\label{sec:interpr-tuned-models}

Among the three graphs depicted in Fig. \ref{fig:c-u-g}, only the
function $\overline c(x)$ has an obvious interpretation as the fitness
wave profile. Using the results obtained in
Ref.~\cite{halla-nelson-TPB-2007}, one can also give an intuitive
interpretation of the functions $u_*(x)$ and $g(x)\equiv u_* \overline
c$. Both functions relate to the phenomenon of fixation. Imagine
sampling an individual at position $x$ and labeling it with an
inheritable label (neutral mutation). As the dynamics proceeds, the
abundance of this label will change due to number fluctuations and the
fitness of its carriers. Eventually, this label will either go
extinct, or become \emph{fixed} in the population. The latter case
occurs if the descendants of the labeled individual take over the
population.

Fixation events are much more likely if the initially labeled
individual belongs to the fitter part of the population. We thus
expect the probability of fixation to be a steeply increasing function
of $x$, similar to the function $u_*(x)$. Indeed, the interpretation
of $u_*(x)$ is precisely that of a fixation probability of a particle
at position $x$, which is derived in the Supporting Information. It is
interesting to note about equation~(\ref{eq:12}) determining $u_*(x)$,
that a very similar equation is well-known to describe the survival
probability of an \emph{unconstrained} branching random
walk~\cite{Derrida:2007p1381}. The only difference is the pre-factor
of $2$ instead of $1$ in the nonlinearity of equation (\ref{eq:12}).

The product $g(x)\equiv u_* \overline c$ also has an intuitive
interpretation in terms of a probability density\footnote{ Note that,
  as required for a probability density, the integral of $g(x) \equiv
  u_* \overline c$ over $x$ is equal to $1$ by virtue of the
  constraint equation (\ref{eq:7}).}. It represents the positional
distribution function of the individual whose descendants eventually
will take over the population~\cite{halla-nelson-TPB-2007}. Even
though there must surely be such a ``lucky'' individual at any time,
it cannot be described deterministically, simply because the fixation
event depends on random future events. Thus, the position of the
``common ancestor of future generations'' can only be described
probabilistically, similar to the position of quantum mechanical
particle.

\begin{figure}
  \includegraphics[width=.4\textwidth]{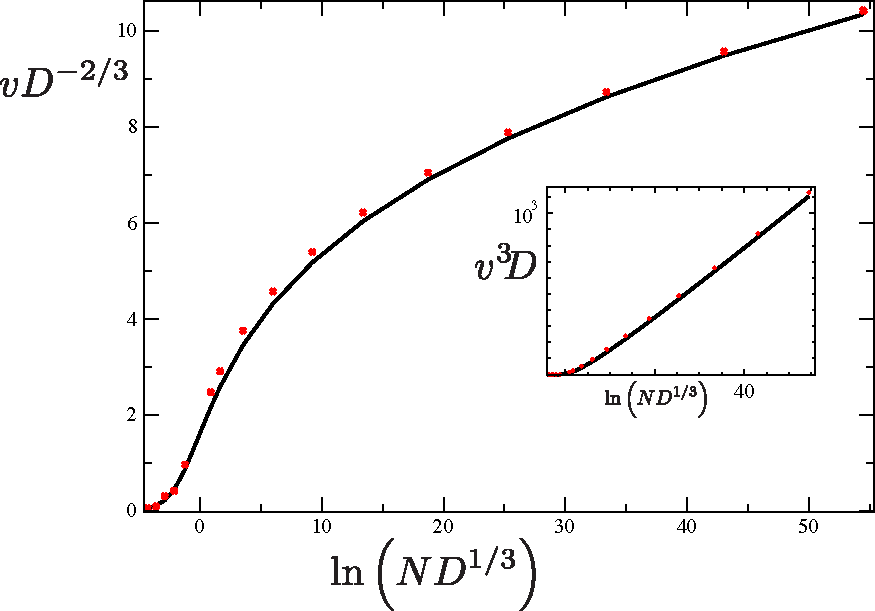}
 \caption{\label{fig:v-N} The relation between the scaled speed, $v
    D^{-2/3}$, and the scaled population size $\overline N D^{1/3}$ as
    obtained numerically for the tuned model (black line), which is
    solvable, and from stochastic simulations of the original
    evolution model~\cite{Tsimring:1996p13322} with a rigid population
    size constraint (red points). Both curves approach each other in
    the limit of large and small $N$. The data are consistent with the
    asymptotic scaling $v\sim \ln^{1/3}N$ (see inset) and $v\sim N$
    for large and small population sizes, respectively.}
\end{figure}

\subsection{Wave speed}
\label{sec:wave-speed}

Next, we compare the evolution model with its tuned counterpart,
considering at first the relation between speed and population
size. From Fig. \ref{fig:v-N}, it can be seen that deviations are
significant only for intermediate values of the control parameter $v
D^{-2/3}$, and disappear in the limit of large and small values. The
scaling $v\sim\ln^{1/3}N$ for large speeds has been observed
previously, but could only be modeled by introducing a cutoff into the
mean-field equations~\cite{Tsimring:1996p13322,Cohen:2005p13342}. Our
tuned model, which exhibits the same scaling, can be used to
rationalize and refine the heuristic cutoff idea. To this end,
consider the closed equation for the mean population density at steady
state,
\begin{equation}
  \label{eq:C-eqn}
   0=D \partial_x^2 \overline c + v \partial_x \overline c +
  (x-x_0)\overline c -\frac{2\overline c^2 e^{v x}}{\int_{x'} \overline
    c^2 e^{v x'}} \;,
\end{equation}
which is obtained by eliminating $u_*$ in equation~(\ref{eq:22}) using
equation~(\ref{eq:24}) and the constraint~(\ref{eq:7}).  The closed
moment equation (\ref{eq:C-eqn}) has the form of
the deterministic approximation except for the last term, which acts
(for large population densities) similar to a cutoff: It is negligible
for small $x$, and completely dominates for sufficiently large $x$
because of the exponential ``amplification'' factor in the
numerator. This leads to an asymptotic wave profile $\sim e^{-v x}$
in the tip of the wave, just as found from an heuristic cutoff in the
reaction rate~\cite{Brunet:VelocityShiftCuttoff}. The location of the
cutoff can be determined by balancing the last term in
equation~(\ref{eq:C-eqn}) with the reaction term $(x-x_0)\overline
c$. In Fig.~\ref{fig:c-u-g}, this crossover point can be identified as
the x-value where $u$ crosses over from exponential to linear.  Notice
that the functional form of the cutoff is quite different from a
step-function, which has been used as a cutoff in
Ref.~\cite{Tsimring:1996p13322} and most subsequent studies. The
peculiar nonlinear form of the cutoff turns out to be essential for
determining the leading order corrections to the wave speed (and its
fluctuations) in the limit of large speeds, as we show in the
Supporting Information. In the opposite limit of small speeds $v\to
0$, we find that the tuned model exhibits $v\sim N/2$, which can be
derived from a perturbation analysis of the moment
equation~\cite{hallatschek2009fisher}. Surprisingly, the exact same
asymptotic is measured for the original evolution model, suggesting
universal behavior not only for large but also for small speeds.

\subsection{Fluctuations}
\label{sec:fluctuations}

Another telling comparison concerns the fluctuations of waves, rather
then the mean density field. At first sight, fluctuations in both
models seem to be very different because they arise from different
statistical ensembles: Whereas wave speeds fluctuate in the evolution
model by Tsimring et al.~\cite{Tsimring:1996p13322}, they are
perfectly constant in the tuned model. On the other hand, population
sizes are constant in the evolution model but fluctuate in the tuned
model. One might expect, however, that both ensembles become
equivalent in the universal large population size limit, similar to
different ensembles in the thermodynamic limit of statistical
mechanics. On the basis of this equivalence hypothesis, one can try to
infer the fluctuations in wave speed of the evolution model (and thus
wave diffusivities) based on the fluctuations of the mortality rate
$\lambda$ in the solvable tuned model (Supporting Information). This
reproduces correctly the known diffusivity scaling
$D_{\textnormal{\scriptsize wave}}\sim \ln^{-3}N$ of Fisher-Kolmogorov
waves~\cite{Brunet:2006p13263} and predicts a novel scaling
$D_{\textnormal{\scriptsize wave}}\sim \ln^{-1}N$ for evolutionary
waves.  We would like to stress, however, that these analytic results
rest on the equivalence of the two descriptions in the large
population size limit, which remains to be proven.

\section{Discussion}
\label{sec:discussion}


As a case study for noisy traveling waves in general, our analysis
concentrated on simple, yet unsolved, models of evolution. These
models effectively describe branching random walks subject to a global
constraint~\cite{Derrida:2007p1381}: The branching random walk has the
effect of modeling the growth of the population and the mutations
through diffusion along the fitness axis. The global constraint fixes
the population size accounting for the fact that natural population
must stay finite due to resource limitations.

Even these simplest evolution models could not be solved previously
because of the nonlinearity associated with the rigid
constraint. However, while a limit on the population size is natural,
there is no obvious biological reason for demanding a non-fluctuating
population size. After abandoning the idealized fixed population size
constraint, we could show that a solvable evolution model with a
closed moment equation can be obtained when the form of the
constrained is tailored to satisfy equation (\ref{eq:12}). The
solvable model was found to reproduce the properties of the fixed
population size model very well, in particular for large and small
population sizes (or wave speeds).

Our approach of model tuning not only applies to evolutionary waves
but to any solitary wave that is generated by a branching random walk
under a global constraint, or some other nonlinearity, to keep
particle numbers finite. Such models, which we may term constrained
branching random walks, share universal features that only depend on
the form of the linear part of the model, as has been found in
extensive computer simulations~\cite{saarloos03}. The universality
class is defined by the linear part of the model, which describes a
branching process and a random walk.  For instance, the important
class of Fisher-Kolmogorov waves~\cite{Fisher:FisherWave,RefWorks:24}
has a reaction term that saturates at a constant value in the tip of
the wave, in contrast to the linear reaction rate gradient that
characterizes the above evolutionary waves. This difference results in
a different scaling of the wave speed and its dependence on number
fluctuations, which is correctly reproduced by our approach beyond the
leading order ``cutoff''
approximation~\cite{Brunet:VelocityShiftCuttoff} (Supporting
Information). Wave speeds are also strongly dependent on the
stochastic particle motion, which is diffusive only in the simplest
cases.  The spread of epidemic waves in the modern human population,
for instance, is dominated by the scale free air transportation
network~\cite{brockmann2006scaling}. This leads to
super-diffusion~\cite{brockmann2007front}, which can be incorporated
in our framework by the use of a fractional diffusion operator in
$\mathcal L$. Apart from these obvious extensions of our approach, our
method can also be applied to problems with time-dependent scenarios
$\mathcal L(t)$ (Supporting Information).  This allows to account for
dynamic driving forces and to study, for instance, waves in temporally
changing environments.

In summary, we have developed and applied an exact method to determine
the universal dynamics of noisy traveling waves. By tuning nonlinear
details of traveling wave models without changing their universal
features, we could close the hierarchy of moment equations for a large
class of noisy traveling waves. These solvable models are amenable to
a simple probabilistic interpretation in terms of fixation
processes. For large population sizes (or weak noise), our description
resembles mean field theory supplemented with a cutoff, and thus
provides, to our knowledge, the first rationale for the heuristic
cutoff
approach~\cite{Tsimring:1996p13322,Brunet:VelocityShiftCuttoff}. The
peculiar, non-local form of the nonlinearity sets the precise location
of the cutoff, which is crucial for the correct prediction of the wave
speed and its fluctuations. More generally, the analysis presented in
this article is applicable to those reaction-diffusion problems in
which a global nonlinear interaction suffices to keep particle numbers
finite. This encompasses simple models of microbial evolution and
range expansions (Fisher-Kolmogorov waves), for which we explicitly
demonstrated the utility of our approach. It will be interesting to
see whether model tuning can be extended to situations where
interactions are inherently local, such as in the problem of directed
percolation~\cite{odor2004universality}.


\begin{acknowledgments}
  We are grateful to many useful discussions with Daniel S. Fisher. We
  thank Richard Neher and Kirill Korolev for stimulating discussions
  and comments on the manuscript.
\end{acknowledgments}



\end{article}

\end{document}